\documentclass[%
 aapm,footnoteinbib,
 mph,%
 amsmath,amssymb,longbibliography,reprint%
]{revtex4-1}

\usepackage{graphicx}
\usepackage{dcolumn}
\usepackage{bm}
\usepackage{amsmath,amssymb}

\usepackage{hyperref}
\usepackage[]{natbib}

\usepackage{tikz}
\usepackage{lipsum}

\begin{document}

\title{Energy relaxation in a vacuum capacitor-resistor circuit: measurement of multiple decays with divergent time constants}


\author{Frank V. Kowalski}%
\email{fkowalsk@mines.edu}

\affiliation{Physics Department, Colorado
School of Mines, Golden CO. 80401 U.S.A.}

\begin{abstract}
The decay of the electrical energy in a resistor-vacuum capacitor circuit is shown to involve multiple relaxation processes, with dramatically different time constants. This is measured using a vacuum capacitor to eliminate the effect of a dielectric between the plates (polypropylene capacitors are shown to exhibit similar behavior). A simple phenomenological model accounts for this behavior in spite of the difficulty in applying Maxwell's equations to such a circuit. These results will lead to a revision of our understanding of the physics of circuits, having particular impact on applications that use capacitors as sensors in collecting precision data (such as found in quantum measurements and dielectric spectroscopy).
\end{abstract}

\date{\today}

 
 \maketitle

\section{\label{sec:level1} Introduction}


Energy redistribution is intrinsic to all dynamical systems. The energy of interest here is that stored in a capacitor while its redistribution involves conversion of electrical to thermal energy in a resistor-capacitor circuit (although the results presented below have implications for energy redistribution via radiative loss, conversion of electrical to mechanical energy, or energy redistribution among other circuit components).

Capacitors are essential lumped-element or inherent components of every electronic system or device, including biological cell membranes. \cite{scott} Their fidelity as sensors impacts the viability of scientific models. The electro-mechanical coupling that causes motion of the plates in a capacitor is used to probe foundational issues in quantum mechanics.\cite{didier,teufel} In addition, some quantum computers utilize capacitors as fundamental components.\cite{qbit,qbit2} Energy relaxation is crucial to the function of these devices.


The capacitor's response is typically modeled with Kirchhoff's current and voltage laws. When applied in steady state these predict the ``ideal'' behavior associated with equal and opposite charges on the capacitor plates and exponential decay in an RC circuit. The same current then flows into one terminal of the capacitor and out of the other. 

However, non-ideal behavior is found in a practical device. For example, non-exponential decay was first studied during the 1850’s using Leyden jars and modeled with stretched exponential functions.\cite{kohlrausch} This response is attributed to the dielectric. Circuit simulation software accounts for this behavior by using a lumped circuit model that is comprised of ideal inductors, resistors, and capacitors. This non-ideal behavior is also used to probe the microscopic structure of the dielectric, a method widely applied in material science and chemistry, referred to as dielectric spectroscopy. \cite{kramer,jonscher,feldman} Modeling of dielectric capacitor behavior using macroscopic variables continues to be of interest. \cite{westerlund,2022,2023}

These effects are mitigated when using a vacuum capacitor. However, the remaining deviation from ideal behavior is caused by the equivalent series inductance that typically produces self-resonance frequencies above $100$ MHz. \cite{murata} The non ideal behavior illustrated here occurs at frequencies smaller than this by at least two and at most eight orders magnitude.

To accurately model decay in a vacuum capacitor one must turn to Maxwell's equations. However, only a handful of solutions that involve steady state behavior have been addressed analytically. \cite{muller} These require a surface charge on the circuit wires to maintain the current as described theoretically \cite{sommerfeld, heald, chabay} and utilized experimentally (where the surface charge functions as a Hall probe). \cite{schade} Applying Maxwell's equations numerically to the relaxation of an RC circuit yields an initial non-exponential decay due to transit time effects. \cite{preyer} However, this calculation does not predict the other decays that are documented below. 

The effect of such surface charge on transient behavior is illustrated by a kink in a circuit wire. Let the current into the kink be greater than that out,  ``then the charge piles up at the ``knee,'' and this produces a field aiming away at the kink. The field opposes the current flowing in (slows it down) and promotes the current flowing out (speeding it up) until these currents are equal, at which point there is no further accumulation of charge and equilibrium is established.'' \cite{griffiths2,moreau}  Such a description is also appropriate for the data presented below which demonstrate unequal currents flowing into and out of the vacuum capacitor.   

There are few first principles analytical models of transient behavior in circuits, even obtaining numerical solutions is challenging. \cite{klee,muller,heald} The objective here is to provide a phenomenological model for the observed novel circuit behavior and relate that to a microscopic description of this energy relaxation process.

\section{\label{sec:level2} Results}

\begin{figure}[ht]
{%
  \includegraphics[width=.9 \columnwidth]{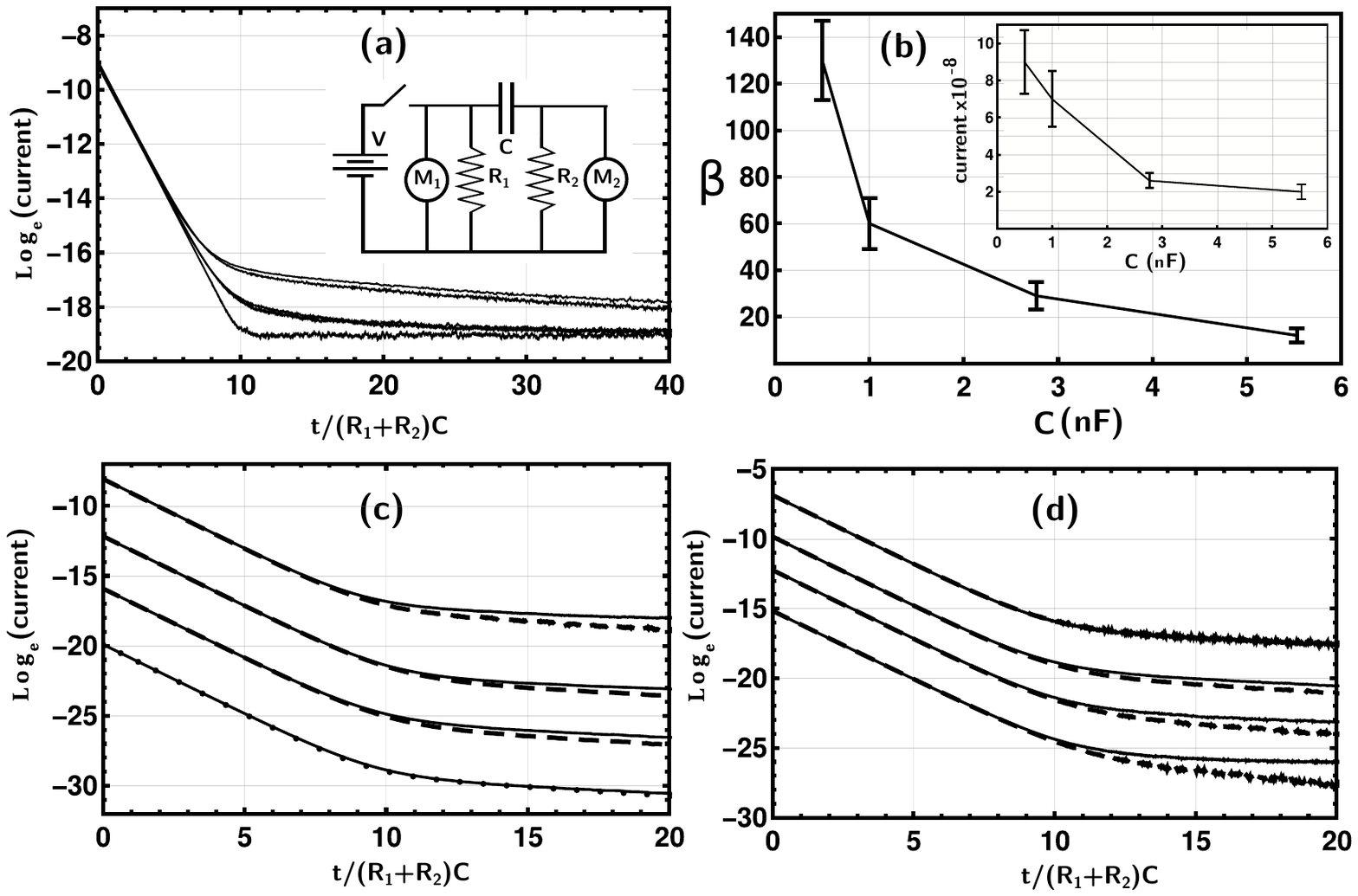}%
}\hfill
\caption{Data for the circuit in the inset of frame (a). The upper lines in frame (a) illustrate currents through $R_{1}$ while the lowest line corresponds to the currents through $R_{2}$ for different vacuum capacitor values while $R_{1}=R_{2}=337.8~{\textrm k}\Omega$. Frames (c) and (d) show currents through $R_{1}$ (solid lines) and  $R_{2}$ (dashed lines) using polypropylene capacitors.  In (c) the three line pairs are for fixed $R_{1}=R_{2}=121~{\textrm k}\Omega$ while the capacitance varies from top to bottom. The lower two pairs of lines are offset along the vertical axis for clarity. The dots in the lowest trace of frame (c) are a fit using a sum of two exponential model to the $2012$ nF $R_{1}$ data. The parameters of this model for the vacuum capacitor are shown in frame (b) where the line connecting the data is only used for clarity. In (d) the capacitance is fixed at $2012$ nF while the resistance increases for the line pairs from top to bottom. The lower three line pairs have been offset from the upper pair along the vertical axis to prevent overlap. A $98$ V power supply was used to energize the circuit. }
\label{fig1}
\end{figure}

Relaxation plots of current vs. time in time constant units are shown in figs.\ref{fig1} (a) for vacuum capacitors and in figs.\ref{fig1} (c) and (d) for polypropylene capacitors. The circuit used is in the inset of fig. \ref{fig1} (a). $M_{1}$ and $M_{2}$ represent the meters that measure the voltages and by Ohm's law these currents.

Fig.\ref{fig1} (a) uses $R_{1}=R_{2}=337.8~{\textrm k}\Omega$ while the capacitance varies. The data for the current through $R_{1}$ in this frame are given by the upper two traces that almost overlap and have C$=0.5177$ nF and C$=1.0354$ nF respectively (the circuit capacitance is constructed from a single $0.5177$ nF capacitor and a parallel combination of two single $0.5177$ nF capacitors). Below these are two $R_{1}$ current data traces that overlap and have C$=2.789$ nF and C$=5.578$ nF respectively (constructed from a single $2.789$ nF capacitor and a parallel combination of two single $2.789$ nF capacitors). All the $R_{2}$ current data for fig.\ref{fig1} (a) overlap forming the lowest trace in this fig. (demonstrating single RC exponential decay down to the noise baseline).

The three pairs of solid and dashed lines in fig. \ref{fig1} (c) illustrate the decay current through $R_{1}=121.5~\textrm{k}\Omega$ (solid line) and the decay current through $R_{2}=121.8~\textrm{k}\Omega$ (dashed line) for the circuit shown in the inset of  fig. \ref{fig1} (a). The resistance is constant while the polypropylene capacitor values of $6.947$ nF, $475.0$ nF, and $2012$ nF increase from top to bottom. The two lower line pairs are offset along the vertical axis to prevent overlap. Fig. \ref{fig1} (d) shows the data for a fixed capacitance of $2012$ nF while $R_{1}=R_{2}=47.0~{\textrm k}\Omega, 121~{\textrm k}\Omega, 180~\textrm{ k}\Omega,~\textrm {and}~432~\textrm {k}\Omega$ increases from top to bottom. All line pairs except the pair for  $R_{1}=R_{2}=47.0~\textrm{k}\Omega$ are offset along the vertical axis to prevent overlap.

\begin{figure}[ht]
{%
  \includegraphics[width=.9 \columnwidth]{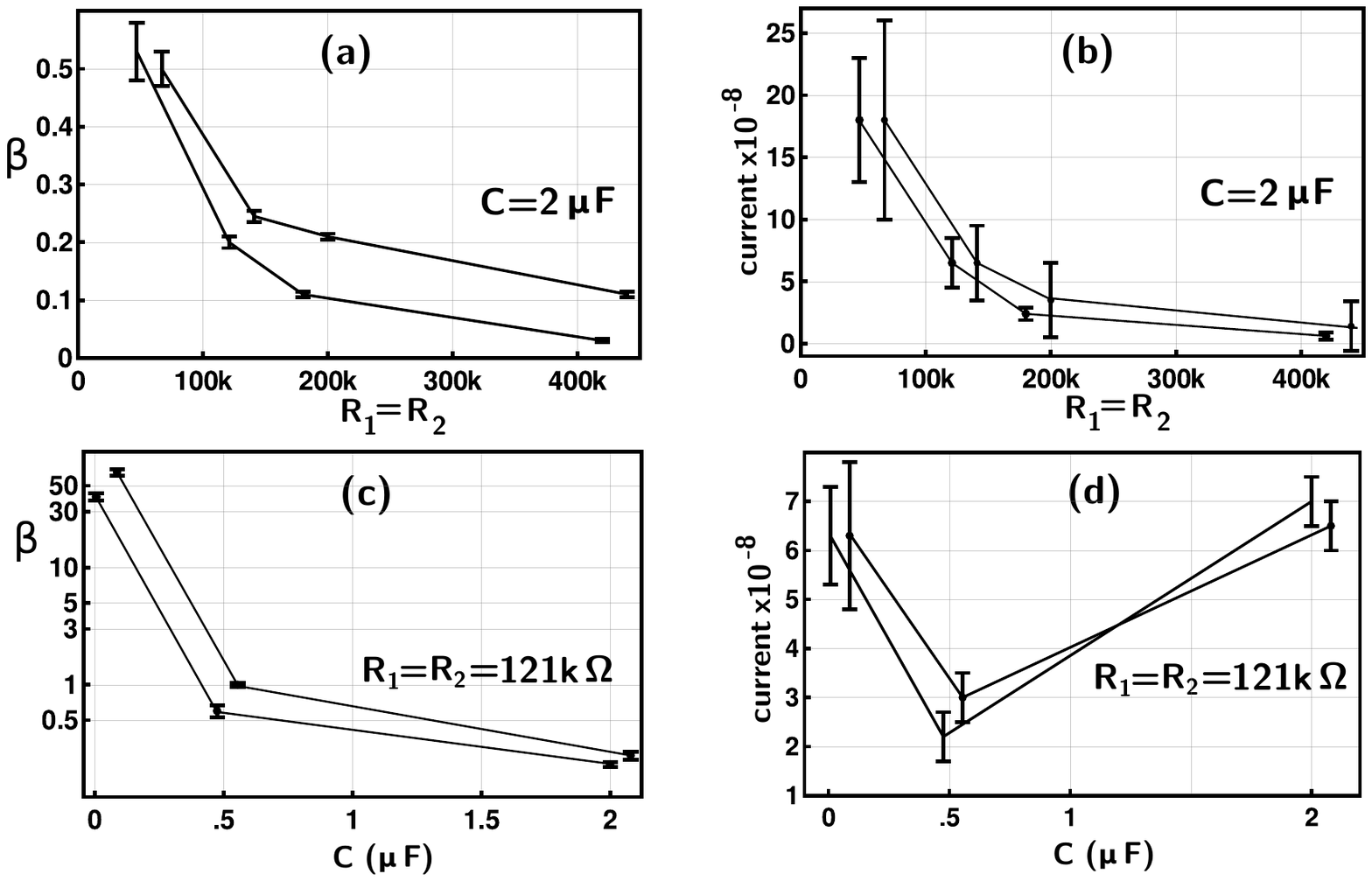}%
}\hfill
\caption{The decay data from figs. \ref{fig1} (c) and \ref{fig1} (d) are fit with a sum of two exponentials model. The decay constant of the smaller exponential term is shown in frames (a) and (c) while the coefficient of this exponential is given in frames (b) and (d). Frames (c) and (d) use the fig. 1(c) data while frames (a) and (b) use the fig. 1(d) data. To prevent overlap of the data in these figures the data associated with $R_{2}$ are offset to the right of the $R_{1}$ data along the horizontal axes in all frames. The line connecting the data points is only used for clarity. }
\label{fig2}
\end{figure}

The lowest line in fig. \ref{fig1} (c) is a reproduction of the $R_{1}=R_{2}=121~\textrm{k}\Omega$ data using the $2012$ nF polypropylene capacitor. The filled circles that overlap with this line (characteristic of a typical fit of the data) are generated from a sum of two exponentials model, the first term of which corresponds to the expected exponential decay for an RC circuit. The parameters of the second term are shown in fig. \ref{fig2} and described below. These parameters for the vacuum capacitor are shown in fig. \ref{fig1} (b). Careful examination of this fit, particularly for longer times (not shown), demonstrate that the two exponential model is inadequate. However, for the majority of the data presented here this model is supported by the data within error.

The unequal currents into and out of the capacitors shown in fig. \ref{fig1} result in a net charge on the capacitor. However, the initial charge on each capacitor plate is not measured. In analogy with the net charge on a hairpin wire this net charge plays a role in the decay behavior.

\section{\label{sec:level3} Phenomenological Model}

A model for the fig. \ref{fig1} data is given by $I=I_{10} \exp[-\alpha t]+I_{20} \exp[-\beta t]$. This is referred to as the sum of exponentials model. For all of the fits to the data presented here $\alpha=1/\textrm{RC}$, where the values of R$=R_{1}+R_{2}$ and C are determined from measurements of the individual circuit components.

The parameters of this model associated with the data of fig. \ref{fig1} (c) and (d) are shown in fig. \ref{fig2}. Fig. \ref{fig1}(c), which varies C for fixed $R_{1}=121~\textrm{k}\Omega$, has the expected model parameters $I_{10}=0.405$ mA and $\alpha = 1650, 1180, 17.30, 4.133$ (from smaller to larger values of C) while the values of $I_{20}$ are shown in fig. \ref{fig2}(d) and the values of $\beta$ are shown in fig. \ref{fig2}(c). Fig. \ref{fig1}(d), which varies R for fixed C$=2012$ nF has the expected model parameters $I_{10}= 1.042~\textrm{mA}, 0.4049~\textrm{mA}, 0.2722~\textrm{mA}, 0.1166~\textrm{mA}$ and $\alpha = 10.63, 4.132, 2.777, 1.190$ (from smaller to larger values of R) while the values of $I_{20}$ are shown in fig. \ref{fig2}(b) and the values of $\beta$ are shown in fig. \ref{fig2}(a).

The model parameters for current flowing through $R_{1}$ in the circuit of fig. \ref{fig1} (a) using a fixed vacuum capacitor with varying resistance is similar to that shown in fig. \ref{fig2} (a) and (b). The similarity in the decay behavior of the vacuum and polypropylene capacitors is manifest in these model parameters as documented in figs. \ref{fig1} (c) and \ref{fig2}. This suggests that the decay mechanisms observed are intrinsic to capacitors.

Plots of current vs time in time constant units allow data to be presented over a range of RC values. However, the parameters of the sum of two exponentials model cannot be inferred from such graphs and must be fit to the current vs time data. 

\begin{figure}[ht]
\includegraphics[width= 1. \columnwidth]{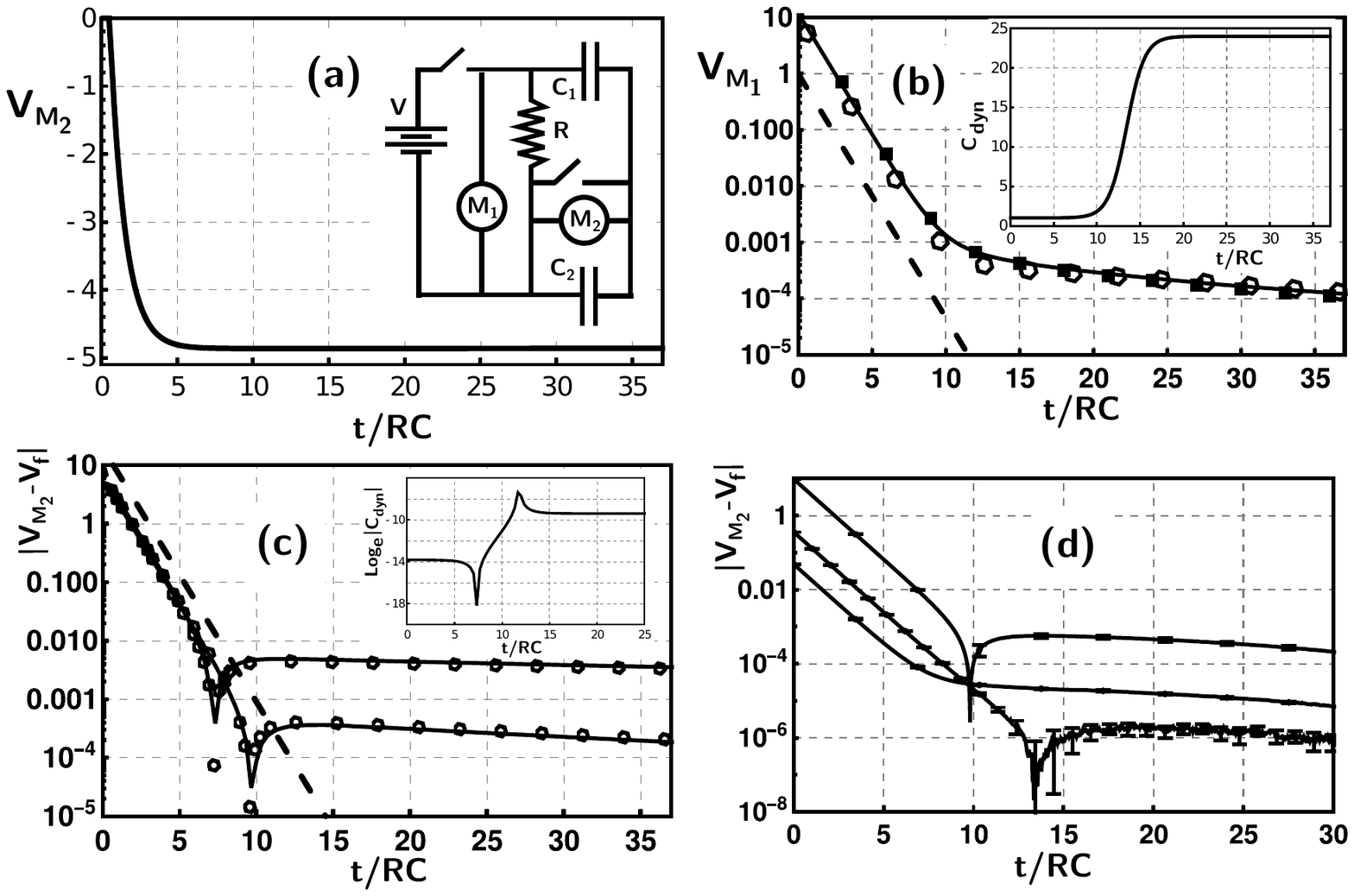} \hfill
\caption{Frames (a), (c), and (d) show meter $M_{2}$ voltage data for the circuit in the inset of frame (a). Voltage data for meter $M_{1}$ is shown in frame (b). The insets in frames (b) and (c) show $C_{dyn}$ for the respective models of the $M_{2}$ data (in microfarads for (b)) while (d) shows the response for different initial voltages. The capacitors are each $2.01~\mu F$. For frame (a) through (c) $R=59.5~\mathrm{k}\Omega$ with $V=9.8$. For frame (d)  $R=578~\mathrm{k}\Omega$ with $V=20, 0.74$, and $0.088$ with a microvolt noise baseline and error bars displayed only for a few of the data points. }
\label{fig3}
\end{figure}

To measure the decay as the charge distributes over an isolated conductor the single capacitor in fig. \ref{fig1} is replaced with two in series, as shown in the inset of fig \ref{fig3} (a) with switches that operate synchronously (within a millisecond while the time constant is $60$ milliseconds). When the switches are closed the isolated right plates of polypropylene capacitors $C_{1}=2.012~\mu F$ and $C_{2}=2.014~\mu F$ have charge and no charge, respectively. When the switches open the right plates then have a initial net charge which is distributed between them during the decay. Without a switch across $C_{2}$ the net charge on the right plates of $C_{1}$ and $C_{2}$ is initially zero (they are grounded before data collection commences). The response of the circuit for these two cases is shown in figs. \ref{fig3} (b) and (c). The latter demonstrates enhancement in the non-exponential decay due to this net charge on the right plates of $C_{1}$ and $C_{2}$.

The sum of exponentials model for the voltage across $C_{2}$, shown in fig. \ref{fig3}, is  $V_{M_{2}}=V^{in}_{f} (1-\exp[-\alpha t]) + V^{out}_{f} (1- \exp[-\beta t])$ with $V^{in}_{f}$ negative. The data is presented in fig. \ref{fig3} as $|V_{M_{2}}-V_{f}|$ in semilog plots where $V_{f}=V^{in}_{f}+V^{out}_{f}$ is approximated by the voltage at $t=100 \mathrm{RC}$.

When the switch across $C_{2}$ is operational the data are obtained as follows. The capacitors are first grounded. Then the switches in fig. \ref{fig3} (a) are closed. The switches are opened and data is recorded for a duration of $t_{f}=100 \mathrm{RC}$. The dips in fig. \ref{fig3} (c) occur when $V=V_{f}$ while the magnitude of the voltage across $C_{2}$ reaches a maximum value and then decreases shortly thereafter. 

Data when the switch across $C_{2}$ is always open is represented in frame (b) by the solid squares while in frame (c) by the lower trace when $t>15~\mathrm{RC}$. Data when this switch is operational is given by the solid lines in frame (b), the upper trace in frame (c) when $t>15~\mathrm{RC}$, and by all traces in frame (d). Time is expressed in time constant units, $t/\mathrm{RC}$, where $C$ is the effective capacitance of $C_{1}$ and $C_{2}$ in series. The diagonal dashed line represents exponential decay with time constant $\mathrm{RC}$. It is offset along the ordinate for clarity. 

The open pentagons and filled squares are sum of exponential model predictions. For fig. \ref{fig3} (b) $V=V_{10} \exp[-\alpha t]+V_{20} \exp[-\beta t]$ where $\alpha = 16.7$, $\beta = 0.7$, $V_{10}=10.0$, and $V_{20}=6.0 \times 10^{-4}$. The tail of the data in fig. \ref{fig3} (c) (causing its deviation from an exponential) is ubiquitous in RC circuits. \cite{swantek} Note that this tail begins when $dV/dt$ in frame (c) changes sign. 

The model predictions for the lower trace in fig. \ref{fig3} (c), shown as the open hexagons, are given by $V=V_{10} \exp[-\alpha t]-V_{20} \exp[-\beta t]$ where $\alpha = 16.7$, $\beta = 0.5$, $V_{10}=7.0$, and $V_{20}=6.0 \times 10^{-4}$. The model for the upper trace data in fig. \ref{fig3} (c) with the switch operational, shown as the open hexagons, is given by $V=V_{10} \exp[-\alpha t]-V_{20} \exp[-\beta t]$ where $\alpha = 16.7$, $\beta = 0.2$, $V_{10}=7.0$ and $V_{20}=5.0 \times 10^{-3}$.

\section{\label{sec:level4} Dynamic Capacitance Model}

A useful parameter in modeling such behavior is the dynamic capacitance, $C_{dyn}=dQ/dV$. This can be rewritten as $C_{dyn} dV/dt=-V/R$ where $V$ is the voltage across the capacitor and $dQ$ the change in charge on the plate where $V$ is measured. Note that Kirchhoff's law for this RC circuit is obtained from this relationship by the substitution $C_{dyn}\rightarrow C$ (for exponential decay $C_{dyn}= C$). The insets of fig. \ref{fig3} (c) and (d) show the dynamic capacitances, determined from the respective sum of exponentials models, that match the data when the switch across $C_{2}$ is operational (this first order ODE then has a solution that is a sum of exponentials). At the instant when $dV/dt=0$ in frame (c) the value for $C_{dyn}$ is undefined.

\begin{figure}[ht]
\includegraphics[width= 1. \columnwidth]{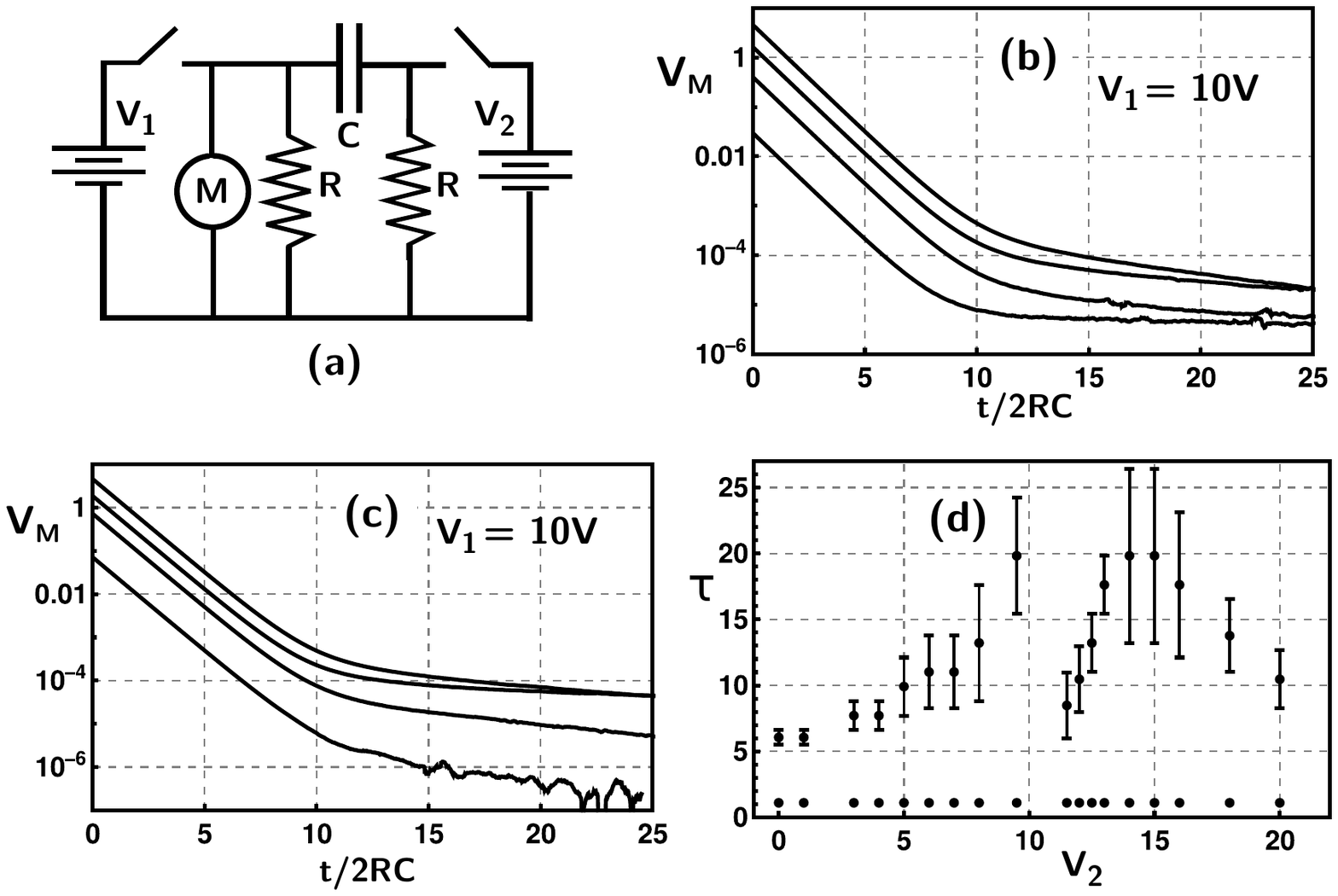} \hfill
\caption{The decay measured for $V_{1}=10$ with different initial voltages on the right side of the polypropylene capacitor for the circuit in frame (a) with $R=274~\mathrm{k}\Omega$ and $C=2.01~\mu$F. The switches are closed to energize the capacitor. They then open simultaneously for the discharge. Frames (b) and (c) illustrate decays for $V_{2}=0, 6, 9, 9.8$ and $V_{2}=20, 14, 11.5, 10.15$ from top to bottom, respectively. The time constants in seconds for the $V_{20} \exp[-\beta t]$ (plotted with error bars) and $V_{10} \exp[-\alpha t]$ (plotted with filled circles) decays are shown in frame (d).  }
\label{fig4}
\end{figure}

\section{\label{sec:level5} Surface Charge Model}

Another modeling approach is to separate the amount of charge on a capacitor plate into two parts: one associated with the energy stored in between the capacitor plates, $Q^{in}$, whose electric field is confined in between the plates, and $Q^{out}$, the charge on the circuit wire and capacitor plate that facilitates the flow of current in the circuit with its electric field not confined. The superposition principle yields a response that is a sum of responses for each type of charge in this separate charge model. Let the $V_{10} \exp[-\alpha t]$ term in the above sum of exponentials model be associated with $Q^{in}$.  Let $Q^{out}_{0}$ be the charge required to direct the current from the power supply into $R$ before the switch is opened. This is expressed as $Q^{out}_{0}=V^{out}_{0} \tau_{out}/ R$, where $\tau_{out}$ is a constant and $V^{out}_{0}$ is the voltage at the plate generated by $Q^{out}_{0}$. For smaller $R$ larger $Q^{out}_{0}$ is required to direct this larger current away from the capacitor plate and into the resistor (similar to the effect that charge on the hairpin wire has on the current flowing through the wire). 

After the switch opens $V^{out}=R dQ^{out}/dt$. The loss of $Q^{out}$ is then determined from $dQ^{out}/dt=-Q^{out}/\tau_{out}$ leading to exponential decay with a time constant independent of $R$ and matching the data in fig. \ref{fig2} (c). The parameters in the model term $V_{20} \exp[-\beta t]$ are $V_{20}=V^{out}_{0}$ and $\beta=1/\tau_{out}$. 

The current due to this second term in the sum of exponentials model for the data in fig. \ref{fig2} is $dQ^{out}/dt= Q^{out}_{0} \exp[-t/\tau_{out}]/\tau_{out}=V^{out}_{0} \exp[-t/\tau_{out}]/ R_{1}$. Fig. \ref{fig2}(d) illustrates this with $V_{20}=V^{out}_{0}/ R_{1}=.13/ R_{1}$. 

For the data in fig. \ref{fig1} each capacitor plate has a different $Q^{out}$ value and therefore a different $V_{20}$ and $\beta$ value to account for the different decays observed from each plate. However, the $V_{10} \exp[-\alpha t]$ term is the same for both plates (with time constant $(R_{1}+R_{2})C$).

Consider the separate charge model applied to the data in fig. \ref{fig3} (d). For decreasing $V$ the initial charge on the right plate of $C_{1}$ decreases. After the switches open this charge flows to the right plate of $C_{2}$, its current limited by $R$. The dip in the response of the upper trace (at $t/\mathrm{RC}=15$) transitions to a response without a dip in the middle trace (at $t/\mathrm{RC}=15$) through essentially a single exponential decay response in the lower trace at ($t/\mathrm{RC}=15$). The voltage across $C_{2}$ is measured to constantly change during the $t=0\rightarrow t_{f}$ time interval for the upper and middle traces; decreasing to minimum value then increasing toward $V_{f}$ for the upper trace and constantly decreasing toward $V_{f}$ for the middle trace. However, the voltage of the lower trace decreases until $t\approx 15 \mathrm{RC}$ and then remains at this value.

This lower trace corresponds to $Q^{out}_{f}\approx 0$.  Larger and smaller $Q^{out}_{f}$ result in positive and negative values of $V_{20}$ with dip and no dip responses as shown in the upper and middle traces. This is supported by fits of this data to the sum of exponential model with parameters $V_{20}/V_{10}=-8.0 \times 10^{-5},-4.3\times 10^{-7}$, and $6.6\times 10^{-4}$ for the upper, lower and middle traces, respectively. With judicious choice of $R$ and $V$ it is therefore possible to generate only the $V=V_{10} \exp[-\alpha t]$ exponential decay across $C_{2}$.

The $\alpha$ decay constant in the first term of the sum of exponentials model is unaffected by such variations in the initial voltage in $V$ while $\beta$ varies. This is evidence for the model that separates charges into $Q^{in}$ and $Q^{out}$. Although not shown in this figure, $V_{M_{1}}$, for the different initial voltages shown in fig. \ref{fig3}(d), varies in essentially the same manner as that shown in fig. \ref{fig3}(b), indicating different values of  $Q^{out}$ for left and right plates of $C_{1}$.

Different values of $Q^{out}_{0}$ on each capacitor plate are also generated by applying different initial voltages to each plate using the fig. \ref{fig4} (a) circuit. Fixing the initial voltage on the left plate while varying the initial voltage on the right plate results in variation of $\tau=1/\beta$ but not the time constant $2\mathrm{RC}=1/\alpha$ associated with $Q^{in}$ in the sum of exponentials model (which is additional evidence for the separation of charges model). Although not shown, the magnitude of $V_{20}/V_{10}$ for these data vary in a similar manner to $\tau=1/\beta$ in fig. \ref{fig4}(d) ranging in value from $1.2 \times 10^{-4}$ to $2.5 \times 10^{-4}$. Evidence that inductance has no impact on these results is given in fig. \ref{fig3}(d) and fig. \ref{fig4} where only a variation of the initial voltage generates a variety of sum of exponentials behavior while the circuit geometry (and therefore the inductance) remains the same.

\section{\label{sec:level6} Discussion}

The geometry of a circuit increases the difficulty in calculating $Q^{out}$ from Maxwell's equations. \cite{muller} Even constructing a numerical model that predicts only a single exponential decay for an RC circuit from Maxwell's equations is nontrivial. \cite{preyer} Yet the data illustrate similar but simple (a sum of exponentials) behavior for geometrically dissimilar capacitors. An additional issue in solving Maxwell's equations is a potential coupling between these decays. However, currents due to $Q^{out}$ are measured even after $Q^{in}$ has exponentially decayed over tens of time constants. During such times $Q^{in}$ is so small that coupling between these decays should not be significant.

A full understanding requires more effort in modeling and data collection. For example, one might conjecture that $Q^{out}$ influences the response of inductors. The effect of such charges in circuits with energy relaxation determined by radiative rather than thermal mechanisms is also of interest, an example being a superconducting LC circuit. The loss due to conversion of the electrical energy stored in the capacitor into mechanical energy (or into more complicated circuit components) may, in addition, be influenced by $Q^{out}$.

The simplicity of the Kirchhoff exponential decay model and its ability to match well the initial decay data are factors that have allowed the results presented above to have been overlooked. Another reason is that the deviations from exponential decay, documented above, are small. An additional reason involves the ubiquitous use of dielectric capacitors. The novel behavior described above might then be attributed to complicated relaxation mechanisms in the dielectric rather than to a fundamental characteristic of a capacitor. A revision of circuit models to account for such behavior is therefore needed, particularly in applications that use a capacitor as a sensor and those that require precision data, such as is found in dielectric spectroscopy and quantum measurements.

\section{\label{sec:level6b} Conclusions}

The decay of the electrical energy in resistor-vacuum capacitor and resistor-polypropylene capacitor circuits involves multiple relaxation processes with divergent time constants and net charge on the capacitor. This novel behavior differs fundamentally from the single exponential decay predicted by circuit simulation software. A microscopic understanding of the sum of exponentials model, that the data support, involves charge on the surface of the circuit components.

\begin{acknowledgments}
I wish to thank Justin L. Swantek, Tony D'Esposito, and Jacob Brannum for useful discussions. 
\end{acknowledgments}

\section*{Data Availability}
The data that support the findings of this study are available from the corresponding author upon reasonable request.


 \hspace{20 mm}

{\bf \label{sec:level7} Appendix 1: Methodology}

A Keysight 34465A digital voltmeter was used to collect data with input impedances of $10^{7}~\Omega$ for the data in fig. \ref{fig1} and greater than $10^{9}~\Omega$  for the data in fig. \ref{fig3}. At least fifteen samples were taken for the vacuum capacitor data and collected at a $50$ kHz rate. The data was process by first taking a mean value of these samples. Next a moving average of four to twenty points was performed on the logarithm of the mean of these samples. One power line cycle integration during the data acquisition was applied to the longer time constant data ($2 \mu$F polypropylene capacitor circuits).

A mechanical switch coupled and decoupled the power supply from the circuit (with an open switch capacitance of less than $0.1$ pF). The circuit layout was similar to that shown in fig. \ref{fig1}. The wire diameter was $0.5$ mm. All the circuits used $1/4$ W $1\%$ metal film resistors (inductance $< 2~\textrm{nH}$). Comet vacuum capacitors models CFMN-2800BAC/8-DE-G and CFMN-500AAC/12-DE-G were used (inductance $< 6~\textrm{nH}$). The circuit was enclosed in a shielded box.

The small $7$ nF polypropylene capacitor allowed the area enclosed by the circuit to vary from $3 \times 10^{-4}$ to $3 \times 10^{-2}$ $\textrm{m}^{2}$. However, the decay data shown in fig. \ref{fig1} (b) did not then change noticeably, demonstrating that inductance is not an important parameter in these results. The similarity in response of a circuit where $dI/dt$ varies by five orders of magnitude is additional evidence against inductance being a important variable. 

\hspace{20 mm}

{\bf \label{sec:level9} Appendix 2: Frequency Domain Data}

\begin{figure}[ht!]
\includegraphics[width= .8 \columnwidth]{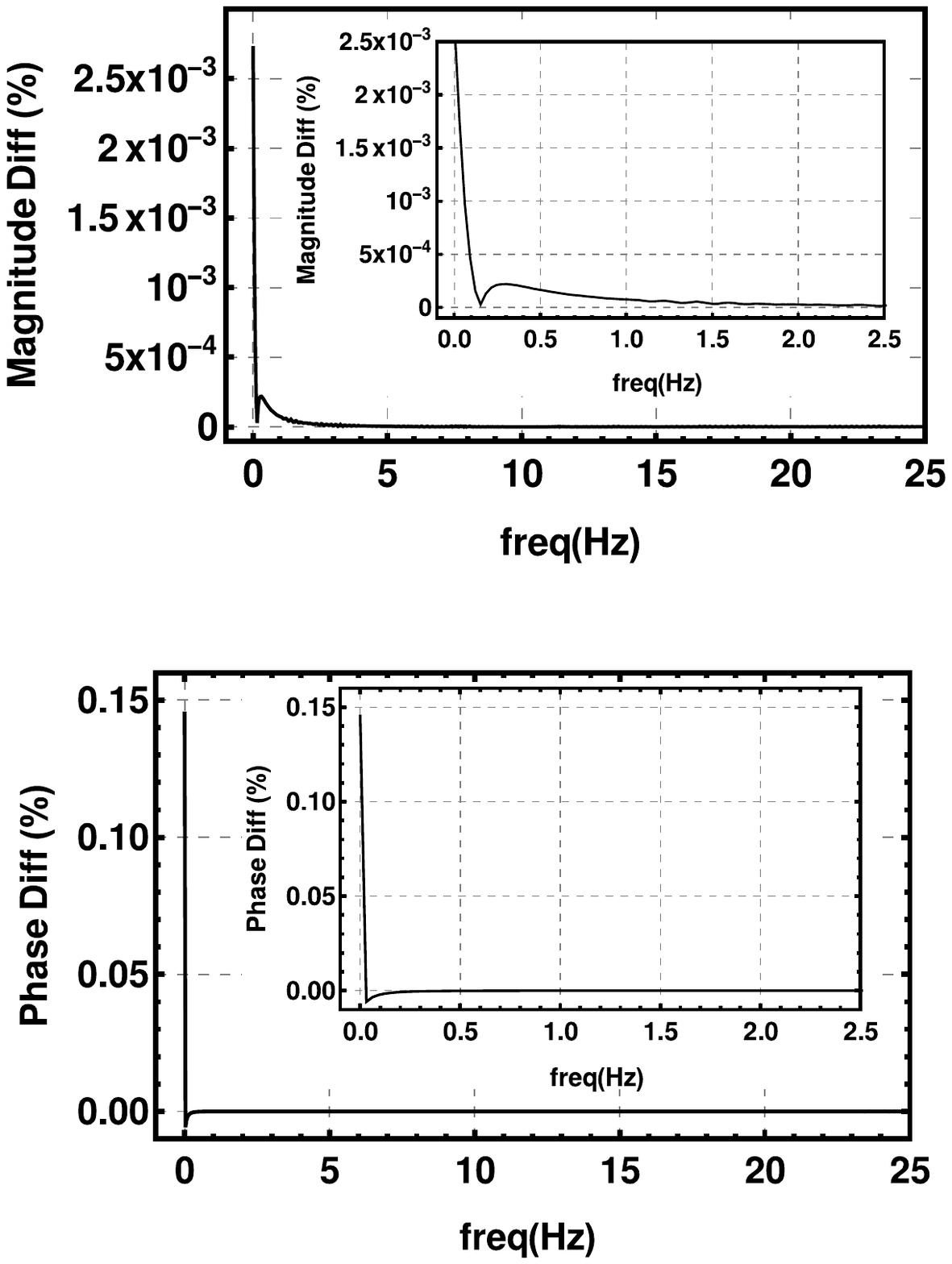} \hfill
\caption{Frequency domain results are shown for the circuit in figs. 3b. Percent differences of the current between the data and exponential decay are given. The upper and lower frames are magnitude and phase graphs.  }
\label{appendixB2}
\end{figure}

The above circuit behavior can also be expressed in the frequency domain. However, the decays with longer time constants predominately manifest at the low end of the spectrum. Since the exponential decay spectrum is largest at these low frequencies it is difficult to resolve on a Bode plot the small differences between the decays with longer time constants and that of the single exponential decay determined by the R and C values. On the other hand, in the time domain the decays with longer time constants occur on a zero background since the initial exponential decay has essentially vanished.

To better illustrate the real and ideal behavior in the frequency domain the difference between the currents in the real and ideal circuits is first determined as a function of frequency. This difference when divided by the ideal current at each frequency yields a percent difference. The magnitude and phase of these percent differences are presented in fig. \ref{appendixB2} for the circuit used in fig.\ref{fig3}.

\clearpage


\providecommand{\noopsort}[1]{}\providecommand{\singleletter}[1]{#1}%

\end{document}